\documentclass[12pt]{article}

\usepackage{amsmath}
\usepackage{amssymb}

\usepackage{indentfirst}
\usepackage{amssymb}
\usepackage{amsfonts}
\usepackage{amscd}
\usepackage{amsbsy}
\usepackage{amsthm}
\usepackage{latexsym}
\usepackage{graphicx,color} 
\usepackage[dvipsnames]{xcolor}
\usepackage[colorlinks]{hyperref}
\hypersetup{linkcolor=blue,citecolor=blue,urlcolor=blue}

\usepackage[symbol]{footmisc} 

\def\nn{\nonumber}       
\def\beq{\begin{eqnarray}}
\def\eeq{\end{eqnarray}}

\def\al{\alpha}
\def\be{\beta}

\def\ga{\gamma}
\def\de{\delta}

\def\ep{\epsilon}

\def\la{\lambda}
\def\na{\nabla}
\def\pa{\partial}

\def\si{\sigma}
\def\om{\omega}
\def\ph{\varphi}

\begin{document}

\begin{center}
\renewcommand*{\thefootnote}{\fnsymbol{footnote}} 
{\Large 
On the vector conformal models in an arbitrary dimension}
\vskip 6mm

{\bf Manuel Asorey}$^{a}$
\hspace{-1mm}\footnote{E-mail address: \ asorey@unizar.es},
\ {\bf Les\l{}aw Rachwa\l{}}$^{b}$
\hspace{-1mm}\footnote{E-mail address: \ grzerach@gmail.com},
\ {\bf Ilya L. Shapiro}$^{b}$
\hspace{-1mm}\footnote{E-mail address: \ ilyashapiro2003@ufjf.br},
\ {\bf Wagno Cesar e Silva}$^{b}$
\hspace{-1mm}\footnote{E-mail address: \ wagnorion@gmail.com}
\vskip 6mm

$^{a}$ Centro de Astropart\'{\i}culas y F\'{\i}sica de Altas Energ\'{\i}as, Departamento de F\'{\i}sica Te\'orica,
Universidad de Zaragoza, E-50009 Zaragoza, Spain

$^{b}$ Departamento de F\'{\i}sica, ICE, Universidade Federal de Juiz de Fora,
\\
Juiz de Fora, 36036-900, Minas Gerais, Brazil
\end{center}
\vskip 2mm
\vskip 2mm


\begin{abstract}

\noindent
The conventional model of the gauge vector field is invariant under
the local conformal symmetry only in the four-dimensional space
($4d$). Conformal generalization to an arbitrary dimension $d$ is
impossible even for the free theory, differently from scalar and fermion
fields. We discuss how to overcome this restriction and eventually
construct four vector conformal actions. One of these models is the
particular case of the previously known conformal theory of $n$-forms
and others are new, up to our knowledge.  In some of these models
the gauge invariance is preserved, two of the new models are
described by local actions with auxiliary compensating scalar fields,
and the extended version of one of these models is on shell equivalent
to the last, non-analytic, purely metric version.
\vskip 3mm


\end{abstract}

\setcounter{footnote}{0} 
\renewcommand*{\thefootnote}{\arabic{footnote}} 
\section{Introduction}
\label{sec1}

It is common wisdom that conformal symmetry plays a very
prominent role in both classical and semiclassical gravity theories.
In the last case, the conformal anomaly \cite{CapDuf-74,duff77}
includes
a classification of the possible terms satisfying the conformal
Noether identities \cite{ddi,DeserSchwimmer}. One can say that
understanding conformal anomaly is the critically important
issue in the semiclassical theory because it is in the heart of the
most important applications in cosmology and black hole physics
(see, e.g., \cite{duff94,PoImpo} for reviews).

On the other hand, many interesting questions are still
unanswered. This concerns, in the first place, the universality
of signs in the anomaly in $4d$ (see, e.g., the recent discussion in
\cite{PoImpo} and \cite{OUP}). It is known that the signs of the
beta functions of the square of the Weyl tensor and the
Gauss-Bonnet term are $({}+\,+\,+\,+{})$ and  $({}-\,-\,-\,-{})$
for the conformal
scalar, fermion, gauge vector, and (we can invoke quantum gravity
too, at this point) conformal quantum gravity. However, for the
``next generation'' of conformal models, i.e., for the
four-derivative scalar \cite{FrTs-superconf,Paneitz}
and three-derivative fermion \cite{FrTs-confSUGRA,GBP-3derfer}, the
sign pattern suffers a flip, i.e., we meet  $(--)$ and  $(++)$ in these
two cases. Is all this just a series of accidents, or there is an
unknown general rule behind this?

In the situation, when this question is without an answer, it looks
interesting to gather more examples, and this makes it interesting
to construct new conformal models. Indeed, this is the problem that has
attracted attention for a long time \cite{DeserNepomechie,Branson}.
We can also cite
Refs.~\cite{Erdmenger-97,ErdmengerOsborn94-98,OsbornStergiou2016}
and references therein. The generalization of the known conformal actions to
other dimensions represents part of this general study. The subject
of the present work is the discussion of this problem for the vector
field with two derivatives in the action.

In what follows, we present the discussion of the generalization of
the $4d$ conformal action of the gauge vector field to an arbitrary
dimension $d$. Such a generalization would enable treating the
free Abelian vector field in the same way as scalar and fermion
field, in the analysis of the one-loop renormalization \cite{tmf84}.
For situations with preserved gauge symmetry, one can also construct
conformally invariant gauge-fixing, similarly to what was done in \cite{Eastwood:1985eh,Queva:2015vaa}.

Let us note that vector conformal operators with two derivatives
may constitute
a basis for conformal field theory (CFT) algebras of primary
vector operators coupled to external geometry, including e.g.
non-trivial background  spacetime. Thus, they might give rise
to new CFT's coupled to gravity. In addition, such operators
may have long ranging applications from studies of conformal
anomalies through applications of vector fields in cosmology to
model building for unified theories (GUT) in particle physics.
The vector conformal models may be even useful in computer
graphics, where the conformal methods and vector fields are
heavily used.

Other possible applications include the description of the
renormalization group flows near fixed points in the theories
with vectors coupled to gravity, related to the asymptotic
safety program in gravity,
and in the condensed matter physics, e.g., concerning vector
excitations of graphene curved sheets varying in time, in the
$3d$ case \cite{Iorio:2013ifa}.

Thus, there may be interesting applications of the results presented
below, but we leave the corresponding discussion to possible further
works and now concentrate on the formal aspect of the problem.
The paper is organized as follows. In the next Sec.~\ref{sec2}, one
can find a brief introduction to the problem. Sec.~\ref{sec3} is a
report on a direct search of the $d$-dimensional conformal vector
operator. As a result, we find that such an operator exists, but the
gauge symmetry should be sacrificed, such that the longitudinal mode
of the vector becomes propagating. A surprising detail is that this
propagation is related to the four-derivative Paneitz operator
\cite{Paneitz} in  $4d$.  In the subsequent Sec.~\ref{sec4}, we present two
models of $d$-dimensional conformal vector operators with the gauge
symmetry preserved, but the auxiliary scalars are required for their
construction. In Sec.~\ref{sec5}, we discuss the universal prescription
for constructing non-analytic $d$-dimensional conformal models,
introduced originally in the Appendix of  Ref.~\cite{anomaly-2004}.
It is shown that this model and also its purely gravitational analog
are on shell equivalent to the metric-scalar model of Sec. 4. Finally,
in Sec.~\ref{sec6}, we draw our conclusions.

\section{The global conformal model with vector field}
\label{sec2}

The action of a vector field in four spacetime dimension has the
form
\beq
S_4(A,g) \,=\,
-\,\frac14\!\int\! d^4x\sqrt{-g}\,
F_{\mu\nu}F^{\mu\nu}
\,=\,
\frac{1}{2}\!\int\! d^{4}x\sqrt{-g}
\;A_{\mu}\big(
g^{\mu\nu}\square - \nabla^{\mu}\nabla^{\nu}
- R^{\mu\nu}\big)A_{\nu},
\label{vec-4}
\eeq
where we restricted our attention to the Abelian model for the sake
of simplicity. When dealing with the generalization to arbitrary $d$,
the case of a non-Abelian field should be considered separately.

Under the local conformal transformation,
\beq
A_\mu\,=\,{\bar A}_\mu,
\qquad
g_{\mu\nu}\,=\,e^{2\si}\,{\bar g}_{\mu\nu} ,
\qquad
\si=\si(x),
\label{conf-4}
\eeq
this action remains invariant, i.e. $S_4(A,g) = S_4({\bar A}, {\bar g}) $.
Our purpose is to formulate the generalization of the transformation
(\ref{conf-4}), that would provide the invariance of the $d$-dimensional
version of the action (\ref{vec-4}).

Direct generalization of (\ref{vec-4}) leads to the functional
\beq
S_d(A,g) \,=\,
-\,\frac14\int d^dx\sqrt{-g}\,
F_{\mu\nu}F^{\mu\nu}
\,=\,
-\,\frac14\!\int\! d^dx\sqrt{-g}\,
g^{\mu\rho}g^{\nu\si}\,F_{\mu\nu}F_{\rho\si}.
\label{vec-n}
\eeq

Assuming that the metric still transforms as in
(\ref{conf-4})\footnote{Let us note that this does not reduce
generality.}, the problem reduces to whether there exists some
real number $w$ and/or a modification in the transformation
of $A_\mu$,
leaving the action (\ref{vec-n}) invariant under
\beq
A_\mu\,=\,e^{w\si}\,{\bar A}_\mu ,
\qquad
g_{\mu\nu}\,=\,e^{2\si}\, {\bar g}_{\mu\nu}.
\label{conf-n}
\eeq
The simplest version concerns global transformation, with
$\si={\rm const}$. Direct replacement of (\ref{conf-n})
into (\ref{vec-n}) shows that the symmetry is achieved for
\beq
w\,=\,\frac{4-d}{2}.
\label{kn}
\eeq

The case of a local transformation is more complicated.
It is easy to check that for $\si\neq{\rm const}$ the transformation
of $F_{\mu\nu}$ does not have the desired form
\beq
F_{\mu\nu} \,=\,\na_\mu A_\nu-\na_\nu A_\mu
\,=\,\pa_\mu A_\nu-\pa_\nu A_\mu
\,\neq \, e^{w\si}{\bar F}_{\mu\nu},
\label{Fn}
\eeq
which is a necessary condition for the global conformal symmetry of
(\ref{vec-n}) (being the prerequisite of the local symmetry). In this formula,
${\bar F}_{\mu\nu} = \pa_\mu {\bar A}_\nu - \pa_\nu {\bar A}_\mu$
and ${\bar A}_\mu$ is from (\ref{conf-n}). Thus, the invariance
under local conformal symmetry requires changing the form of the
action and of the transformation rule for $A_\mu$.

\section{Construction of the vector conformal operator}  
\label{sec3}

As a first step, we consider the direct construction of a vector
conformal operator. In this section, we shall derive the conformal
model known from the work by Deser and Nepomechie
\cite{DeserNepomechie}. Let us note that when the first version
of the present paper was prepared, we were not, unfortunately, aware
of this well-known paper. However, we decided to preserve this
section in the subsequent version for the sake of generality and
also owing to some new details described in what follows.

Our starting point is the two-derivative action quadratic
in the vector  field $V_\mu$ and without self-interactions, whose
basis consists of a general combination of possible scalar invariants,
that can be  constructed using a covariant vector field $V_{\mu}$ on
a general spacetime background. The corresponding action is
defined as
\beq
S\,=\,
-\frac{1}{2}\!\int\! d^{d}x\sqrt{-g}\;V_{\mu}O^{\mu\nu}V_{\nu},
\label{action1}	
\eeq
where the operator of the energy dimension two has the form
\beq
O^{\mu\nu}
\,=\,
a_{1}\nabla^{\mu}\nabla^{\nu}+a_{2}g^{\mu\nu}\square
+a_{3}g^{\mu\nu}R+a_{4}R^{\mu\nu}
\label{O}
\eeq
and $ a_{i}$ are arbitrary real coefficients. We reserve the
notation $A_{\mu}$ for the gauge field and thus use $V_{\mu}$
here since the gauge symmetry is not demanded, in general.
We assume that the conformal weight of the covariant vector field
$V_{\mu}$ is $w$, such that the conformal transformation is
(cf. with (\ref{conf-n}))
\beq
V_{\mu}
 \,\,\longrightarrow\,\,
\bar{V}_{\mu}=e^{-w\sigma}V_{\mu},\quad
g_{\mu\nu} \,\,\longrightarrow\,\,
\bar{g}_{\mu\nu}
=e^{-2\sigma}g_{\mu\nu}.
\label{transVg}
\eeq
Under the infinitesimal conformal transformation (first 
order in $\si$), the action \eqref{action1} transforms as
\beq
\bar{S} \,=\, S\left(\bar{g}_{\mu\nu},\,\bar{V}_{\mu}\right)
\,=\,S+\delta_{c}S,
\label{CT_law2}	
\eeq
where
\beq
\delta_{c}S
& = &
\frac{1}{2}\!\int\! d^{d}x\sqrt{-g}\;
\bigg\{(d-4+2w)\sigma
\;V_{\mu}O^{\mu\nu}V_{\nu}
+ \big[(d-2+w)a_{1}
\nonumber
\\
&&
- \,
(d-2)a_{4}\big] V_{\mu}(\nabla^{\mu}\nabla^{\nu}\sigma)V_{\nu}
- \big[(1-w)a_{2}+2(d-1)a_{3}+a_{4}\big]
V_{\mu}g^{\mu\nu}(\square\sigma)V_{\nu}
\nonumber
\\
&&
+ \,
\big[(d-2+w) a_{1} + 2a_{2}\big]
V_{\mu}(\nabla^{\nu}\sigma)\nabla^{\mu}V_{\nu}
+ \big[(d-4+2w)a_{2}\big]
V_{\mu}g^{\mu\nu}(\nabla_{\tau}\sigma)\nabla^{\tau}V_{\nu}
\nonumber
\\
&&
- \,
\big[(2-w)a_{1}+2a_{2}\big]
V_{\mu}(\nabla^{\mu}\sigma)\nabla^{\nu}V_{\nu}\bigg\}.
\label{CV_action1}
\eeq
In this and subsequent formulas, the parentheses restrict the action
of covariant derivatives, e.g., $\na A = A\na + (\na A)$.

The conformal invariance requires $ \delta_{c}S=0$, i.e., the
integrand in \eqref{CV_action1} has to vanish. This condition gives
the system of equations for the coefficients
\beq
&& d-4+2w=0, 
\nn
\\
&&
(d-2+w)a_{1}-(d-2)a_{4}=0, 
\nn
\\
&& (1-w)a_{2}+2(d-1)a_{3}+a_{4}=0, 
\nn
\\
&&
(d-2+w)a_{1}+2a_{2}=0, 
\nn
\\
&& (2-w)a_{1}+2a_{2}=0.
\label{12}
\eeq
Note that some equations are degenerate for the dimensions
$d=1$ and $d=2$. Let us first look at these special dimensions.
\begin{itemize}
\item
In the case $d=1$, we get $ a_1=a_2=a_4=0 $. Since, on top of that,
$R=0$, the operator is geometrically irrelevant in $d=1$.
\item For $d=2$ we find $w=1,\;\;a_{1}=a_{2}=0$, and
$a_{3}=-\dfrac{a_{4}}{2} $. Thus,
\beq
O^{\mu\nu}\big|_{d=2}
\,=\,
a_{4}\left.\Big(R^{\mu\nu}-\frac{1}{2}g^{\mu\nu}R\Big)\right|_{d=2}
\,=\, 0. 
\label{op_2d}
\eeq
\end{itemize}
Thus, in dimensions $d=1$ and $d=2$, the conformal vector
operator of the given type does not exist.
In other dimensions, it is easy to see that the system of equations
in (\ref{12}) can be
solved with $w$ from (\ref{kn}) and with the following coefficients:
\beq
a_{2}=-\frac{d}{4}a_{1},
\quad
a_{3}=\frac{(d-4)d^{2}}{16(d-2)(d-1)}a_{1}
\quad
\textrm{and}
\quad
a_{4}=\frac{d}{2(d-2)}a_{1}.
\label{relations_1}	
\eeq
The resulting conformal operator is uniquely defined, up to the
overall arbitrary real coefficient $a_1$,
i.e.,\footnote{This operator was
originally found in \cite{DeserNepomechie}.
Furthermore, it is the particular $n =2$ case of the conformal
operator for the $n$-forms of gauge field strength as described in
Refs.~\cite{Erdmenger-97,OsbornStergiou2016}.}
\beq
O^{\mu\nu}
\,=\,
a_{1}
\Big[\nabla^{\mu}\nabla^{\nu}-\frac{d}{4}g^{\mu\nu}\square
+ \frac{(d-4)d^{2}}{16(d-2)(d-1)}g^{\mu\nu}R
+ \frac{d}{2(d-2)}R^{\mu\nu}\Big].
\label{Opd}
\eeq
As it should be expected, this expression becomes singular at
$d\to 1$ or $d\to 2$. Until the end of this section, we do not
consider the special dimensions $d=1,2$. On the other hand, in $d=4$, the
coefficient $a_3=0$ and we arrive at the well-known operator
for the Maxwell field on a general background,
\beq
O^{\mu\nu}\big|_{d=4}
\,=\,
a_{1}\big(\nabla^{\mu}\nabla^{\nu}
- g^{\mu\nu}\square
+ R^{\mu\nu}\big).
\label{op_4d}	
\eeq
Choosing $a_1=1$, one obtains the action (\ref{vec-4}).
It is remarkable that it is possible to recover the action of
the electromagnetic field theory in $d=4$, using only the requirement of
the local conformal invariance in curved space, i.e., without
demanding the gauge symmetry. Let us stress that this feature is
typical only for this special dimension.

Coming back to the case of the general dimension $d$, direct
calculations (we skip
the details in this part since they are cumbersome) demonstrate
that the action \eqref{action1} is also invariant under the finite
conformal transformations (\ref{transVg}). One can note that the
operator in \eqref{Opd} is unique and that the conformal invariance
does not concern surface terms.

Since we derived the action (\ref{action1}) with (\ref{Opd}) without
demanding gauge invariance, the next step is to check out how it
behaves under the Abelian gauge transformation,  i.e.,
\beq
V_{\mu}\rightarrow V'_{\mu}=V_{\mu}+\nabla_{\mu}f ,
\label{gatrans}
\eeq
where $ f=f(x) $ is an arbitrary scalar field.

Performing the transformation (\ref{gatrans}), after some integrations
by parts, and using the contracted Bianchi identity, we find that the
conformal action transforms as
\beq
\label{gauge_transf}
&&
S'
\,=\,
S\,-\,\frac{1}{2}a_{1}\!\int\! d^dx\sqrt{-g}\;\bigg\{
\frac{\left(d-4\right)}{2}\nabla_{\nu}V_{\mu}
\Big[
g^{\mu\nu}\square-\frac{d^{2}}{4(d-2)(d-1)}g^{\mu\nu}R
\nonumber
\\
&&
\qquad
+ \,\frac{d}{(d-2)}R^{\mu\nu}\Big]f
+ \frac{d(d-4)}{8(d-1)}V_{\mu} (\nabla^\mu R) f
+ \frac{d-4}{4}f\,\Big[\square^2
+ \frac{d}{d-2}R^{\mu\nu}\nabla_{\mu}\nabla_{\nu}
\nonumber
\\
&&
\qquad
-\, \frac{d^{2}}{4(d-2)(d-1)}R\square
+ \frac{d}{4(d-1)} \left(\nabla^{\mu}R\right)\nabla_{\mu}
\Big]
f\bigg\}.
\eeq
Thus, for $d\neq 4$ the action is not gauge invariant.

Since the gauge invariance of the action (\ref{vec-4}) means the
absence of a longitudinal mode, it is worth exploring this mode in
the model (\ref{Opd}). Let us perform a York decomposition of
the vector field, $V_\mu = V_\mu^\perp+V_\mu^\parallel$. Here
the field $V_{\mu}^{\parallel}$ is regarded as a divergence of
a scalar field $V_{\mu}^{\parallel}=\na_\mu\varphi$ and the
transversality conditions reads
$\nabla^{\mu}V_{\mu}^{\perp}=0$.\footnote{We do not intend
to discuss the practical implementation of these requirements on
an arbitrary metric background, but simply suppose that it can be
done.}

In the new variables, the action reads
\beq
&&
S
\,=\,
-\frac{a_1}{2}\!\int\! d^{d}x\sqrt{-g}\;
\bigg\{ V_{\mu}^{\perp}\Big[-\frac{d}{4}g^{\mu\nu}\square
+\frac{d^{2}(d-4)}{16(d-1)(d-2)}g^{\mu\nu}R
+\frac{d}{2(d-2)}R^{\mu\nu}\Big]V_{\nu}^{\perp}
\nonumber
\\
&&
\qquad
+ \,\frac{d-4}{4}\varphi
\Big[
\square^{2}
+ \frac{d}{d-2}R^{\mu\nu}\nabla_{\mu}\nabla_{\nu}
- \frac{d^{2}}{4(d-1)(d-2)}R\square
+\frac{d}{4(d-1)} \left(\nabla^{\mu}R\right)\nabla_{\mu}
\Big]\varphi
\quad
\nonumber
\\
&&
\qquad
+\,\frac{d(d-4)}{2(d-2)}\Big[
\left(\nabla_{\nu}V_{\mu}^{\perp}\right) R^{\mu\nu}\varphi
\,+\,
\frac{d-2}{4(d-1)}V_{\mu}^{\perp}(\nabla^{\mu}R)
\varphi \Big]\bigg\}.
\label{Y-decomp}
\eeq
It is easy to see that the longitudinal mode decouples from the
transverse one only in $d=4$, where it disappears. Thus, we can
conclude that the generalization of the $d=4$ action (\ref{vec-4})
to an arbitrary dimension $d$ makes the longitudinal part of the
vector field propagating. This means we gain the new degree of
freedom compared to the original gauge-invariant action. This is
the price one has to pay for the conformal symmetry in this model.
One can say that this is a kind of a conformal St\"uckelberg
procedure, being, however, quite different from the well-known
one of Deser \cite{Deser70}.

An interesting detail is a similarity between the longitudinal
$\ph\ph$ mode in (\ref{Y-decomp}) and the Paneitz conformal
operator \cite{Paneitz} (see also \cite{Orsted}).
\beq
&&
\Delta_4
\,=\,
\square^2
+ \frac{4}{d-2}R_{\mu\nu}\nabla^{\mu}\nabla^{\nu}
- \frac{d-6}{2(d-1)}(\na_\mu R) \na^\mu
- \frac{d^2 - 4d + 8}{2(d-1)(d-2)}R\square
\nn
\\
&&
\,\,
+\, (d-4)\bigg[
\frac{d^3- 4d^2+16d-16}{16(d-1)^{2}(d-2)^{2}}R^{2}
- \frac{1}{(d-2)^{2}}R_{\mu\nu}R^{\mu\nu}
- \frac{1
}{4(d-1)}(\square R)\bigg].
\label{PanOp}
\eeq
One can note that the $\ph\ph$ terms in the brackets of
\eqref{Y-decomp} coincide with (\ref{PanOp}) in the $d=4$ limit.
This coincidence does not hold in other dimensions, however this
can be seen as a shortcut to the Paneitz operator in $d=4$ (another
link between the two operators was discussed in \cite{Queva:2015vaa}).
The origin of this relation looks unclear, and perhaps can be added
to the list of  open problems concerning conformal operators.
Another interesting aspect is that the
$V_{\mu}^{\perp}\ph$-sector of Eq.~(\ref{Y-decomp}) also includes
the coefficient $\frac{d-2}{d(d-1)}$, that is a typical value for the
two-derivative scalar operator \cite{Orsted}.

All considerations presented above address only the free field model
(\ref{action1}). One can consider various  possible extensions including
interaction terms. The addition of these terms may be seen as a kind
of analogy with the non-Abelian gauge symmetry, but for just one
vector field. The first option is the term $(V^{\mu}V_{\mu})^{m}$.
This interaction term in $d=4$ and, respectively for $m=2$, was
introduced in Eq.~(8.122) of  \cite{book} for the axial vector field
related to torsion  (see also related discussion of the conformal
transformations with torsion in \cite{BuchShap85} and \cite{torsi}).
The $d$-dimensional action with such a term has the
form\footnote{In the dimensions $d\geqslant 6$ one can construct
more conformal interactions, e.g., using the products of
$V^{\mu}V_{\mu}$ and $(\na_\al V_\be - \na_\be V_\al)^2$.}
\beq
S_{2}=-\frac{1}{2}\!\int\! d^{d}x\sqrt{-g}\;
\Big\{ V_{\mu}O^{\mu\nu}V_{\nu}
+ \lambda(V^{\mu}V_{\mu})^{m}\Big\},
\label{action2}	
\eeq
where $ \lambda $ is a coupling constant. Taking the infinitesimal
conformal variation of the action \eqref{action2} and requiring that
$\delta_{c}S_{2}=0$, we arrive at the value of $m$,
\beq
m = \frac{d}{d-2},
\eeq
independent of the coupling $\la$. The conditions for the quadratic
part of the action are the same as in \eqref{relations_1}, with
$w= 2-d/2$.

\section{Conformal models with auxiliary scalars}
\label{sec4}

In this section, we consider a few simpler ways to generalize
the action (\ref{vec-4}) into an arbitrary dimension $d$, preserving
the invariance under the local conformal symmetry.

\subsection{Extended connection}
\label{sub-41}

One possible solution of the problem is based on the modification
of the definition of $F_{\mu\nu}$ in (\ref{Fn}). This solution is
rooted in the similar approach used to explore the conformal
transformations in the models of gravity with torsion (see, e.g.,
\cite{torsi} and references therein, pioneer work \cite{Deser70},
and also subsequent papers with a similar procedure, e.g.,
\cite{conf,Shocom,asta} and \cite{Percacci2011,Percacci2012}).
We assume that the covariant derivative of the vector field is
constructed with a modified affine connection,
\beq
D_\mu A_\nu = \na_\mu A_\nu - K^\la_{\,\,\,\nu\mu}  A_\la,
\qquad
K^\la_{\,\,\,\mu\nu}
= \frac12\big(\de^\la_\mu \pa_\nu\ph - \de^\la_\nu \pa_\mu \ph\big),
\label{Kaff}
\eeq
where $\ph$ is an additional (can be called auxiliary) scalar field.
We assume that the transformation rule for this field has the form
\beq
\ph\,=\,{\bar \ph} + \ga \si,
\label{phtrans}
\eeq
where $\ga$ is the parameter to be found from the modified version of
the {\it l.h.s.} of (\ref{Fn}), with a new version of the field tensor
${\cal F}_{\mu\nu}$,
based on a new covariant derivative $D_\mu$. Equation (\ref{phtrans}) is
called to supplement the transformations of the metric and gauge
vector field (\ref{conf-n}).

Substituting (\ref{Kaff}) into the new definition of the field tensor
\beq
{\mathcal F}_{\mu\nu} \,=\,D_\mu A_\nu - D_\nu A_\mu
\,=\,\pa_\mu A_\nu-\pa_\nu A_\mu + A_\nu \pa_\mu \ph
- A_\mu \pa_\nu \ph,
\label{tilFn}
\eeq
after a small algebra we find that the condition
\beq
{\mathcal F}_{\mu\nu}
\,=\, e^{w\si}{\bar {\mathcal F}}_{\mu\nu}
\label{newcond}
\eeq
is satisfied for
\beq
\ga\,=\,-\,w\,=\,\frac{d-4}{2}.
\label{gamn}
\eeq

Finally, the conformally symmetric action has the form
\beq
\tilde{S}_d(A,g) \,=\,
-\,\frac14\!\int\! d^dx\sqrt{-g}\,
{\mathcal F}_{\mu\nu}{\mathcal F}^{\mu\nu},
\label{conf-vec-n}
\eeq
that coincides with (\ref{vec-4}) in the limit $d=4$ and $\varphi=0$.
One of the
main features of (\ref{tilFn}) and (\ref{conf-vec-n}) is that these
two objects do not have the gauge symmetry, at least not under the
usual gauge transformation. One can, of course, try to look for
the modified form of this transformation. However, this issue
was explored in the models with torsion \cite{novello,rr}, hence
we skip this part.

Thus, we arrive at the new form of the vector field action,
remaining conformally invariant in an arbitrary dimension $d$. The
price to pay is the $\mathcal{O}(d-4)$-modification of the
affine connection for a scalar field and the
$\mathcal{O}(d-4)$-violation of gauge invariance. Indeed,
both issues do not contradict the scheme of the proof of
conformal one-loop renormalizability, given in \cite{OUP}
(and more complete one in \cite{tmf84}).

Furthermore, it remains unclear how to make a generalization
for a non-Abelian vector field, due to the non-linearity in the
field tensor
\ $G^a_{\mu\nu} \,=\,\pa_\mu A^a_\nu - \pa_\nu A^a_\mu
+ g f^{abc} A^b_\mu A^c_\nu$ and the respective
inhomogeneity of $G^a_{\mu\nu}$ with respect to $A^a_\mu$.

\subsection{Simpler way of using scalar field}
\label{sub-42}

Consider an alternative way of providing conformal symmetry in the
modified version of the Abelian vector field action (\ref{vec-4}).
In this case, we do not change the
transformation rule for the vector field in (\ref{conf-4}), but
instead insert the auxiliary scalar in the action. In this case, there
is no problem with the gauge invariance and with the non-Abelian
version of the model. Thus, we directly consider the $d$-dimensional
action
\beq
S^{\ast}_d(A,g) \,=\,
-\,\frac14\!\int\! d^dx\sqrt{-g}\,\,\Phi^\la\,
G^a_{\mu\nu}G^a_{\rho\si}\,g^{\mu\rho}g^{\nu\si},
\label{vec-NN}
\eeq
where $\la = \frac{4-d}{2-d}$. Indeed, for the sake of quantum
theory, it is useful to have a free part of the vector action, that
enables one to construct propagator and separate vertices. To
get this, let us set $\Phi = \chi_0+\chi$, where $\chi_0$ is a
constant of the mass dimension $(d-2)$, and regard $\chi$ to be
a new scalar field. To achieve the conformal invariance of
(\ref{vec-NN}), the transformation law for the new scalars should be
\beq
\Phi \,=\, {\bar \Phi} e^{(2-d)\si}
\qquad
\mbox{or}
\qquad
\chi_0 + \chi \,=\, (\chi_0+ {\bar \chi}) e^{(2-d)\si}\,.
\label{trans}
\eeq
With these definitions, the combination $\sqrt{-g}\,\Phi^\la
\,g^{\mu\rho}g^{\nu\si}$ is conformally invariant, therefore
providing the invariance of the action in \eqref{vec-NN}.

It is worth mentioning that the Faddeev-Popov procedure needs
only a minor change, i.e., the gauge fixing action should be
\beq
S^{\ast}_{{\rm gf},\,d}(A,g)
\,\,=\,\,
-\,\frac{1}{2\om} \!\int\! d^dx\sqrt{-g}\,t^a\,\Phi^\la\,t^a
\,\,=\,\,
\frac12 \!\int\! d^dx\sqrt{-g}\,t^a\,Y_{ab}\,t^b,
\qquad
t^a = \na_\mu A^{a\,\mu}\,.
\label{gfact}
\eeq
Thus, the change concerns only the weight operator $\,Y_{ab}\,$
and not the gauge conditions. The new weight operator is local and
therefore does not contribute to the divergences. The action of
gauge ghosts constructed in this model has a standard form. It is not
conformally invariant, but this is true even in $d=4$. This fact
does not imply the non-invariance of the one-loop divergences
\cite{tmf84}. In general, since $\la = \mathcal{O}(d-4)$, all
the diagrams with the field $\chi$ come with the factors $(d-4)$,
hence the one-loop divergences are not expected to change.

\section{Non-analytic approach to conformal models}
\label{sec5}

For the sake of completeness, let us mention the existing, albeit
not very much discussed (except the appendix of \cite{anomaly-2004})
method of constructing conformal actions in an arbitrary dimension
$d$. Let us illustrate how this non-analytic approach to construct
$d$-dimensional conformal model works for the massless vector
field. In this case, a single formula can replace many words, so the
desired result is
\beq
S^{\star}_d(A,g) \,=\,
-\,\frac14\!\int\! d^dx\sqrt{-g}\,
\big(G^a_{\mu\nu}G^{a\,\mu\nu}\big)^{\frac{d}{4}},
\label{vec-n-nonloc}
\eeq
that is certainly conformal for any $d$. In the aforementioned
Ref.~\cite{anomaly-2004}, the same idea was used with the
square of the Weyl tensor
$C_{\mu\nu\!\rho\si}C^{\mu\nu\!\rho\si}$ instead of the square
$G^a_{\mu\nu}G^{a\,\mu\nu}$ in Eq.~(\ref{vec-n-nonloc}).

Thinking about the applications to quantum gravity, up to
some extent, this solution looks less interesting and less
useful compared to the three others discussed above. The reason
is that, in this case, it is unclear how to use the Faddeev-Popov
procedure.
As it was mentioned in
\cite{anomaly-2004}, the similar gravitational term also has only
restricted interest since, for example, it cannot give rise to the free
propagation of gravitons around flat spacetime for $d>4$.

On the other hand, the situation may be different if we trade the
action (\ref{vec-n-nonloc}) to the  extended version of the model.
Introducing one more scalar-dependent term in (\ref{vec-NN}),
we get
\beq
S^{\ast}_{d,{\rm ext}}(A,g)
\,=\,
\!\int\! d^dx\sqrt{-g}\,\bigg\{
-\,\frac14\, \,\Phi^\la\,G^2
\,\,+\,\, \frac{\tau}{4}\,\Phi^\al\bigg\},
\label{vec-NN-ext}
\eeq
where $\tau$ is a new coupling constant and we used condensed
notation $G^2=G^a_{\mu\nu}G^a_{\rho\si}
\,g^{\mu\rho}g^{\nu\si}$.
It is easy to check that the local conformal invariance of this
expression requires fixing $\al=\frac{d}{d-2}$. The point is that,
using the equation of motion for $\Phi$, the model  (\ref{vec-NN-ext})
is equivalent to (\ref{vec-n-nonloc}). In detail, the on shell
condition gives
\beq
\Phi \,=\,\Big(\frac{\la}{\al\tau}\,G^2\Big)^{1/(\al-\la)}.
\label{onshell}
\eeq
After some algebra, the on shell action is found in the form
\beq
S^{\ast}_{d,{\rm ext}}(A,g)\Big|
\,=\,
\!\int\! d^dx\sqrt{-g}\,\Big(-\frac{1}{4}+ \frac{d-4}{4d}\Big)
\,\Big(\frac{d-4}{\tau d}\Big)^{\frac{d}{4}-1}\,
\big(G^2\big)^{\frac{d}{4}}.
\label{vec-NN-osh}
\eeq
After an obvious constant reparametrization of the gauge field
$A_\mu^a$ and the gauge coupling $g$, this expression coincides
with (\ref{vec-n-nonloc}). Notice that in the limit $\tau \to 0$ and for $d>4$, the on shell
action (\ref{vec-NN-osh}) is singular, as it should, because when the last term in the action
(\ref{vec-NN-ext}) vanishes there is no equivalence between the two models.

The procedure described above can be used also in the purely
gravitational action \cite{anomaly-2004}, with the Weyl tensor
used instead of the Yang-Mills field tensor,
\beq
S^{\star}_d(g) \,=\,
-\,\frac{1}{2\la}\!\int\! d^dx\sqrt{-g}\,
\big(C_{\mu\nu\!\rho\si}C^{{\mu\nu\!\rho\si}}\big)^{\frac{d}{4}}.
\label{W-d-nonloc}
\eeq
The procedure described above (including the values of $\al$,
$\la$ and other coefficients) remains the same and, as a result,
we arrive at the on shell equivalent representation of the
conformal action.

The Lagrangian in (\ref{vec-n-nonloc}) and similar gravitational
version (\ref{W-d-nonloc}) are non-analytic for odd dimensions.
However, we have shown
that they give rise to classical dynamics which can be described
by the equivalent models which are local and analytic.

The last observation concerns the new conformal model
of the duality-invariant conformal extension of a modified Maxwell’s
theory with self-interactions \cite{Sorokin}. The approach analogous
to (\ref{vec-n-nonloc}) can be certainly applied to generalizing it to
dimensions $d \neq 4$, but this extension may not be unique. It 
would be interesting to find a scalar mapping for both conformal 
models (\ref{vec-n-nonloc}), (\ref{W-d-nonloc}), and for that of 
\cite{Sorokin}, using the procedure described in \cite{fReddi}.
One can also see \cite{Hassaine,LiMiao,JingPanChen} and further
references therein,  for different utilizations of the Power-Maxwell
models, such as (\ref{vec-n-nonloc}).

\section{Conclusions}
\label{sec6}

We considered four different ways to provide the  invariance under
local conformal transformations for the action of massless vector
field in an arbitrary spacetime dimension $d$. These constructions
teach us a few small, but potentially useful lessons. In the first
model, we learned
that the requirement of conformal symmetry in the presence of the
metric leads to the standard action of Abelian vector model in $d=4$,
but not in any other dimension, and without demanding the Abelian gauge
symmetry. For $d = 4 + \ep$, we have found that the violation of
gauge symmetry in the lowest-order $\mathcal{O}(\ep)$-terms
is partially related to the $4d$ scalar Paneitz operator
\cite{FrTs-superconf,Paneitz}.

The next two models show that the desired generalization of
conformal model to $d\neq 4$ is always achieved by means of
an extra scalar field. All these models can be seen as different
versions of the conformal St\"uckelberg procedure. Out of
these models, the (\ref{vec-NN}) is certainly the most
appropriate for the Lagrangian quantization. Furthermore, it can
be shown that its extended version possesses the on shell
equivalence with the non-analytic conformal model described in Sec. 5.

The aforementioned non-analytic model is interesting mostly
by its
universality. It is clear that one can use the procedure qualitatively
similar to (\ref{vec-n-nonloc}) for constructing conformal models
in an arbitrary $d$ starting from a special value $d_0$ where the
initial model is conformal. This procedure is supposed to work
for different field contents and also for different numbers of derivatives.

\section*{Acknowledgments}
Authors are very grateful to Omar Zanusso for pointing to us relevant
references concerning the model discussed in Sec.~\ref{sec3}.
WCS is grateful to CAPES for supporting his Ph.D. project.
LR would like to acknowledge hospitality of the Federal
University of Juiz de Fora and thanks FAPEMIG for a technical
support. The work of M.A. is partially supported by Spanish
MINECO/FEDER grant PGC2018-095328-B-I00 and DGA-FSE
grant 2020-E21-17R.
The work of I.Sh. is partially supported by Conselho Nacional de
Desenvolvimento Cient\'{i}fico e Tecnol\'{o}gico - CNPq under the
grant 303635/2018-5, by Funda\c{c}\~{a}o de Amparo \`a Pesquisa
de Minas Gerais - FAPEMIG under the grant PPM-00604-18, and by
Ministry of Education of Russian Federation under the project
No. FEWF-2020-0003.



\end{document}